\documentclass[11pt,preprint]{aastex}
\usepackage{url}
\def\lta{\lower2pt\hbox{$\buildrel {\scriptstyle <}
   \over {\scriptstyle\sim}$}}
\def\gta{\lower2pt\hbox{$\buildrel {\scriptstyle >}
   \over {\scriptstyle\sim}$}}

\begin{document}

\title{Supernovae-induced accretion and star formation in the inner
   kiloparsec of a gaseous disk}

\author{Pawan Kumar\altaffilmark{1} \& Jarrett L. Johnson\altaffilmark{1,2}}

\altaffiltext{1}{Astronomy Department, University of Texas at Austin,
Austin, TX 78712}

\altaffiltext{2}{Theoretical Modeling of Cosmic Structures Group,
Max-Planck-Institut f{\"u}r extraterrestrische Physik,
Giessenbachstra\ss{}e, 85748 Garching, Germany}

\begin{abstract}
We consider the effects of supernovae (SNe) on accretion and star formation
in a massive gaseous disk in a large primeval galaxy. The gaseous disk we
envisage, roughly 1 kpc in size with $\gta10^8$M$_\odot$ of gas,
could have formed as a result of galaxy mergers where tidal interactions 
removed angular momentum from gas at larger radius and thereby concentrated
it within the central $\sim 1$kpc region. We find that SNe lead to accretion
in the disk at a rate of roughly 0.1--1 $M_\odot$ yr$^{-1}$ and induce star
formation at a rate of $\sim 10$--100 M$_\odot$ per year which contributes 
to the formation of a bulge; a part of the stellar velocity dispersion is
due to SNa shell speed from which stars are formed and a part due to 
the repeated action of stochastic gravitational field of SNe remnant 
network on stars. The rate of SNa in the inner kpc is shown to be self 
regulating, and it cycles through phases of low and high activity.  
The supernova-assisted accretion transports gas from about one kpc to
within a few pc of the center. If this accretion were to continue
down to the central black hole then the resulting ratio of BH mass to the
stellar mass in the bulge would be of order $\sim 10^{-2}$--$10^{-3}$,
in line with the observed Magorrian relation.

\end{abstract}

\keywords{accretion: theory, method: analytical -- supernovae -- stars: formation -- galaxies: bulges}

\section{Introduction}

The CO observations of ultra-luminous infra-red galaxies (ULIGals) find
the gas mass in the inner regions of the galaxy to be about 
5$\times10^9$M$_\odot$, and the average particle density to be of order 
10$^3$ cm$^{-3}$ and the kinetic temperature of molecular gas 
$\sim 50-100$ K (e.g. Downes 
\& Solomon, 1998; see Sanders \& Mirabel, 1996, for a review). The gas
in the central $\sim$kpc region of the galaxy is likely to have come
from distances of the order of 10 kpc when it lost some of its angular momentum 
due to gravitational tidal torques (Barnes \& Hernquist, 1992, and 
references therein; see Barnes, 2002, for a more recent numerical simulation). 

The inner kiloparsec region of most young massive galaxies is likely composed
of a gaseous disk with a mass of several hundred million solar masses, 
ULIGal being at the extreme end of the mass distribution. This gaseous disk is
expected to host star formation at a large rate.
Some of these stars will explode and give rise to shock waves in the gaseous
disk which will spawn both more star formation and accretion of gas toward the center of the galaxy.
We consider these processes analytically in some detail in this paper, paying special attention to
the effect they might have on the evolution of the central parts of the galaxy and on the growth of a
central black hole.

There exists a large body of work on the subject of galaxy
mergers, star formation and black hole growth e.g. Sanders et al. (1988), 
Kauffmann \& Haehnelt (2000), Kawakatu \& Umemura (2002), Granato et al. (2004), Croton et al. (2006),
Kauffmann \& Heckman (2009), Chen et al. (2009), (pl. see Kormendy \& 
Kennicutt, 2004, for a review), and sophisticated numerical simulations
e.g. Barnes \& Hernquist (1991, 1996),
Mihos \& Hernquist (1996), Di Matteo et al. (2005), Springel et al. (2005),
Hopkins et al. (2005), Hopkins \& Hernquist (2008). What is different in 
the present work is that we try to capture some of the basic 
properties of this complex system using analytic results for supernova (SNa)
remnant evolution and other simple physical scalings which are hard to 
capture in numerical simulations due to the large ratio of galaxy size
and SNa shell radius.

The physical system we consider is described in \S2 along with the
effect SNe have on accretion. Bulge formation as a product of 
SN-induced star formation in the gaseous disk is 
discussed in \S3. The main conclusions and uncertainties 
of this study can be found in \S4. 

\section{Supernovae-induced accretion in a gas disk}

Numerical simulations of gaseous disks (e.g. Wada \& Norman, 2001) 
find the medium to be multi-phase and highly filamentary as a result of
star formation and stellar explosion. For the analytical calculations
in this work, where our primary interest is in average disk properties, 
we consider a simplified disk structure that ignores its filamentary density
structure; accordingly, the gas distribution is taken to be a smooth function 
of distance from the center. Many of the results reported in this work, 
as we shall see, have a weak dependence of the interstellar medium (ISM) 
density and therefore the error introduced by the assumption of smooth density 
field in the disk should not be large.

We consider a disk, roughly 1 kpc in radius, consisting of 
$\gta 10^8$ M$_\odot$ in gas that came from larger radius ($\sim 10$s kpc) 
due to e.g., tidal interaction with another galaxy. The mean gas density 
in the disk at 1 kpc is $\sim 10^3$ cm$^{-3}$. Stars form and die in the
disk at a certain rate; we take the SNa rate in the disk to 
be $d f_{sn}(r)/dA$ per unit area per year. 
The interaction between the remnant of a supernova with 
the gaseous disk carves out a cavity in the disk. The gas swept
up by a supernova is compressed into a thin shell during the snowplow phase,
which starts when the thermal cooling time scale of the shocked gas is 
less than the age of the remnant. This means that a certain amount of gas
is pushed closer to the galactic center by the supernova shock wave and, of
course, a certain amount is pushed outward. However, unless gas pushed inwards
by the supernova shock loses angular momentum it will be pushed back
out to a larger radius when the shock weakens. We shall estimate later in
this section the loss of angular momentum for gas pushed closer to the 
galactic center by the SNa, in order to determine if it is sufficient to 
keep the swept up gas at a smaller radius. 
But first, we estimate the amount of gas swept up and pushed to smaller 
radius by a supernova.

Let us consider a supernova going off in the disk at radius $r$. The number 
density of particles in the disk at this radius is $n(r)$,
the vertical scale height is $H(r)$, and the mean rotation 
speed of gas on its circular orbit is $V_{orb}(r)$. During the adiabatic 
expansion of the SNa shell -- the Sedov-Taylor phase -- the
radius, speed and temperature of the shock front are given by
\begin{eqnarray}
R_s(t) &=& 3.1 t_4^{2/5} n_3^{-1/5} E_{51}^{1/5} \;{\rm pc}, \label{st1}
   \\\nonumber\\[-.5cm]
V_s(t) &=& 123 t_4^{-3/5} n_3^{-1/5} E_{51}^{1/5} \;{\rm km}\,
   {\rm s}^{-1},  \label{st2} \\\nonumber\\[-.5cm]
T(t) &=& 2.1\times10^{5} t_4^{-6/5} n_3^{-2/5} E_{51}^{2/5}\;{\rm K}^o,
    \label{st3}  
\label{st}
\end{eqnarray}
where $t_4=t/10^4$yrs, $E_{51}=E/10^{51}$erg, and $n_3=n/10^3$cm$^{-3}$ is 
the particle number density of gas in the disk at radius $r$. The Sedov-Taylor
phase ends when the radiative cooling time $t_{cool} = 0.67 k T/(n\Lambda)$ is
equal to the dynamical time; $\Lambda\approx10^{-16} T^{-1}$ergs cm$^3$ 
s$^{-1}$ (Blondin et al. 1998). The radiative phase begins at time
\begin{equation}
t_{snow} \approx 723 n_3^{-9/17} E_{51}^{4/17}\; {\rm yrs}.
\label{tsnow}
\end{equation}
Subsequently, during the snowplow phase ($t>t_{snow}$), the evolution is 
described by (cf. Chevalier, 1974; eq. 26)
\begin{eqnarray}
R_s(t)  & \approx & 0.8 (t/t_{snow})^{0.31} n_3^{-7/17} E_{51}^{5/17}\;{\rm pc},
      \label{rsnow}  \\\nonumber\\[-.5cm] 
V_s(t)  & \approx & 595 (t/t_{snow})^{-0.69} n_3^{2/17} E_{51}^{1/17}\;{\rm km}
  \,{\rm s}^{-1},  \label{vsnow}  \\\nonumber\\[-.5cm]
M_s(t) & \approx & 133 (t/t_{snow})^{0.93} n_3^{-4/17} E_{51}^{15/17}
   \;{\rm M}_\odot, \label{msnow}  
\end{eqnarray}
where $M_s$ is the mass of gas swept up by the SNa remnant.
These equations are valid only as long as the shell radius is less than 
the vertical scale-height $H$:
\begin{equation}
H = {2^{1/2} (C_s^2+V_t^2)^{1/2}\over \Omega} = {2^{1/2} r (C_s^2+V_t^2)^{1/2}
   \over V_{orb}} \sim (500 {\rm pc})\, r_3^{3/2} \left( { V_t\over
    10 {\rm km\,s}^{-1}} \right) \left( {M(r)\over 10^8 M_\odot}
      \right)^{-1/2}\,
\label{scaleH}
\end{equation}
where $r_3$ = $r$/10$^3$ pc, $C_{\rm s}\sim 1$km s$^{-1}$ is the sound speed, 
$V_t$ is the RMS turbulence velocity in the gaseous disk (produced by 
SNa explosions and winds from early type stars), and $M(r)$ is the 
total mass enclosed within radius $r$.

The Toomre $Q$ parameter for the gravitational instability of the gaseous disk 
is:
\begin{equation}
Q = {2 (C_s^2+V_t^2)^{1/2} \Omega\over \pi G \Sigma} \sim 1 V_{t,6} M_8^{-1/2}
    r_3^{1/2},
\end{equation}
where $\Sigma$ is the mass density per unit area, $V_{t,6}\equiv 
V_t/10^6$cm s$^{-1}$, and $M_8\equiv M(r)/10^8 M_\odot$. We see from
the above equation that the disk is gravitationally unstable, and will 
support an on-going star formation activity.

Ignoring density gradients in the disk ($R_s\ll r$ \& $R_s<H$), the 
expansion of a SNa shell is
nearly spherically symmetric until the Coriolis or centrifugal 
force per unit mass becomes of order the deceleration of
SNa remnant. This occurs when V$_s \Omega_{orb} \approx d$ V$_s/dt$,
 or $t\approx 0.69/\Omega_{orb}$, and defines a characteristic time 
when the shell is no longer spherical.
\begin{equation}
t_{shear} \approx 7\times10^6 r_3 V_{orb,2}^{-1} \; {\rm yrs} \,
\label{tshear}
\end{equation}
where $V_{\rm orb,2}$ = $V_{\rm orb}$ / 100 km~s$^{-1}$.  In fact, at this time the
magnitude of the shear velocity across the shell -- 
$R_s\,d\Omega_{orb}/d\ln r$ -- is of order the speed of the SNa remnant.  The 
SNa shell is thus dispersed by the shear flow in the disk and mixed with 
the ambient ISM on this time scale. 
The radius, velocity and mass of the remnant at this time is 
\begin{eqnarray}
R_s(t_{shear})  & \sim & 14\,  n_3^{-0.25} E_{51}^{0.22} r_3^{0.31} 
   V_{orb,2}^{-0.31} \;
   {\rm pc}, \label{rshear}  \\\nonumber\\[-.5cm]
V_s(t_{shear})  & \sim & 1.1\, n_3^{-0.25}E_{51}^{0.22}V_{orb,2}^{0.69} 
   r_3^{-0.69}
    \;{\rm km}\, {\rm s}^{-1}, \label{vshear} \\\nonumber\\[-.5cm]
M_s(t_{shear})  & \sim & 7\times10^5 \,r_3^{0.93} V_{orb,2}^{-0.93} n_3^{0.25}
    E_{51}^{0.66} \;{\rm M}_\odot.
\label{mshear} 
\end{eqnarray}
Note that $R_s(t_{shear})\lta H$, and therefore the SNa shell is confined
within the disk unless $r\lta 10$pc. Since the shell velocity is small 
compared with the orbital speed when $R_s\sim H$ pc, the swept
up gas in the disk cannot escape the galactic potential. 

During the snowplow phase the swept-up gas is compressed into a thin shell, 
forming a hollow sphere. Therefore, half of the swept up ISM gas, of mass 
$M_s/2$, is pushed closer to the center of the galaxy
by a distance $\sim R_s/2$. For an ensemble of SNe going off in the disk
the net amount of gas that is accreted at $r$ depends on the SNe rate as 
a function of distance from the galactic center ($r$) and is given by
\begin{equation}
\dot{M}_{acc} \sim R_s {d\over dr} \left( {M_s(r)\over2} [2\pi r R_s]
  {d f_{sn}\over dA} \right) \sim M_s (R_s/r)^2 f_{sn}(r)
\label{acc1}
\end{equation}
where $f_{sn}(r)$ is the cumulative SNa rate within the radius $r$. We assumed
that ($M_s R_s f_{sn}/r$) is an increasing function of $r$ in deriving the
second part of the above equation; otherwise, SNe would lead to a net outflow 
of gas in the disk to larger distances. Since $M_s R_s\propto n^{0.01}$ 
during the snowplow phase (combine eqs. \ref{rsnow} \& \ref{msnow}, \&
substituting for $t_{snow}$ from eq. \ref{tsnow}) 
as long as $f_{sn}(r)$ increases with distance faster than $r^{1}$ there is a
net mass accretion even when SNa remnants are spherically symmetric. 
For a Mestel disk the mass enclosed inside radius $r$ increases linearly
with $r$, and in that case there is no net accretion -- for spherical SNe -- 
if the rate of stellar explosions is linearly proportional to gas mass. 
We shall see below that the condition on $f_{sn}(r)$ is relaxed when we 
consider the distortion of SNe remnants by Coriolis and centrifugal forces. 
Equation (\ref{acc1}) for accretion rate is also modified when SNe shells
undergo collision before $t_{shear}$; this is discussed below.
However, in any case we need to make sure that gas pushed inward 
by a SNa loses specific angular momentum; otherwise, it would be 
pushed back out when the shock becomes sufficiently weak.

It should be noted that half of a SNa shell has negative angular momentum
(as seen by an observer at the center of the explosion comoving with the 
disk), and the other half, with prograde velocity field, has positive 
angular momentum. The magnitude of the total positive/negative angular 
momentum grows during the adiabatic expansion as $|L_\pm|\approx M_s V_s r/2 
\propto 1/V_s$ (the second part of this relation follows from energy 
conservation during the adiabatic expansion phase).
However, the angular momentum does not increase much during 
the snowplow phase when $M_s V_s$ is approximately constant. The magnitude
of the negative angular momentum carried inward by the lower half of a 
SNa shell ($|L_-|$) is quite large and that can lead -- as we shall see
shortly -- to an accretion rate in a gas rich disk (like ULIGal)
of order a few solar masses per year. 

It can be shown that, for a freely expanding shell, particles on half of 
the shell with negative angular momentum -- that have retrograde velocity
as seen by a comoving observer at the center of explosion -- 
will on average descend a distance 
of $\Delta r = (8V_{orb}/\pi V_s - 1) R_s^2/4r$ in the radial direction 
(as a result of Coriolis and centrifugal forces in the comoving frame), as
long as $V_s\gta R_s |d \Omega_{orb}/d\ln r|$. Particles on the other half of
the shell with positive angular momentum, or prograde velocity, will move 
outward the same distance. Thus, the shell is continuously distorted with 
time as a result of tidal stretching and Coriolis force.

Part of the lower half of the shell (lying closer to the galactic center)
with transverse velocity component in the direction of the orbital 
velocity has a positive angular momentum with respect to 
the center of explosion, and the part with transverse velocity opposite to 
the orbital motion has negative specific angular momentum. 
The net amount of negative angular momentum carried by the distorted lower 
half of the shell, during the snowplow phase, as seen by an observer 
comoving with the center of explosion, is
\begin{equation}
\delta L \approx \Delta r (L_{-}/R_s) \approx M_s R_s V_{orb}/\pi.
\label{am1}
\end{equation}

When the lower half of the shell is mixed -- due to shear stretching or 
collision with 
other shells -- the net mean specific angular momentum of the mixed fluid
is smaller by $\sim 2 R_s V_{orb}/\pi$ than a particle at the center
of the explosion orbiting the galaxy. This is a sufficient amount of negative 
angular momentum to keep the mixed lower half of shell moving on a circular 
orbit at a smaller radius of $(r-R_s/2)$. 
Thus, the constraint discussed earlier on $f_{sn}$ for inward mass accretion 
is relaxed because there is a net outward transport of positive angular 
momentum associated with each SNa shell.

The rate of angular momentum transported by an ensemble of SNe in the disk is
\begin{equation}
\dot L \sim \delta L (2\pi r R_s) {d f_{sn}\over dA} \sim {2M_s\over \pi} 
   R_s V_{orb} (R_s/r) f_{sn},
\label{am2}
\end{equation}
and the accretion rate resulting from this outward angular momentum transport
is
\begin{equation}
\dot{M}_{acc} \approx {\dot L \over r V_{orb} }\approx {2M_s f_{sn}\over 
    \pi} \left( {R_s\over r}\right)^2, 
\label{acc2}
\end{equation}
which is similar in magnitude to that given by equation (\ref{acc1}), i.e. 
SNe transport mass inward for a larger class of functions for $f_{sn}(r)$ than
suggested by the discussion following equation (\ref{acc1}).

The accretion rate depends on the shell radius ($R_s$) at the time when 
SNe shells collide with each other or when the shell is dispersed and 
mixed due to shear flow in the disk. We consider both of these cases below.

Given a supernova rate of $f_{sn}$ per year within radius $r$ of the disk,
the mean separation between SNe that occurred within time $t$ is
\begin{equation}
d_{sn} \approx 1.8\, \frac{r}{1 {\rm kpc}}\, \left(\frac{t}{1 {\rm yr}}\right)^{-1/2} f_{sn}^{-1/2}\;{\rm kpc}.
\end{equation} 
SNe shells collide when their size is of order the mean separation i.e.
$R_s(t)\approx d_{sn}$. The mean collision time, $t_{coll}$, is estimated 
using equation (\ref{rsnow}), and is given by
\begin{equation}
t_{coll} \approx 1.7\times 10^5\, n_3^{0.31} f_{sn}^{-0.62} E_{51}^{-0.27} 
   r_3^{1.23}\; {\rm yrs}.
\label{tcoll}
\end{equation}
The ratio of the collision time and the time it takes for a SNa shell 
to be dispersed due to shear velocity in the disk is
\begin{equation}
{t_{shear}\over t_{coll} } \approx 40 n_3^{-0.31} f_{sn}^{0.62} r_3^{-0.23} 
     V_{orb,2}^{-1} E_{51}^{0.27}.
\end{equation}

The accretion rate when $t_{shear}/t_{coll}\lta 1$ is given by (using
eqs. \ref{rshear}, \ref{mshear} \& \ref{acc2})
\begin{equation}
\dot{M}_{acc} \sim M_s(t_{shear}) f_{sn} \left[{R_s(t_{shear}) \over r}
   \right]^2 \sim 80 \, f_{sn} n_3^{-0.25} r_3^{-0.45}
     V_{orb,2}^{-1.55} E_{51}^{1.1}\; {\rm M}_\odot\,{\rm yr}^{-1}.
\label{acc3}
\end{equation}
The condition for a remnant not to collide with others before $t_{shear}$
places a limit on the supernova rate of
\begin{equation}
f_{sn} \lta 3\times10^{-3}\, n_3^{1/2} V_{orb,2}^{1.6} r_3^{0.37} 
   E_{51}^{-0.44}\; {\rm yr}^{-1} \equiv f_{sn}^{coll}.
\label{fsn}
\end{equation}
The second part of eq. (\ref{acc3}) is valid only when the SNa rate is
less than the rate given in eq. (\ref{fsn}); the accretion rate 
corresponding to this limiting SNa rate is
\begin{equation}
\dot{M}_{acc} \sim 0.3 n_3^{0.25} E_{51}^{0.66} r_3^{-0.08} 
   V_{orb,2}^{0.05} \;{\rm M}_\odot \;{\rm yr}^{-1},
\label{acc3a}
\end{equation}
The accretion rate has a very weak dependence on $f_{sn}$ when the
SNa rate is larger than the rate given in equation (\ref{fsn}), i.e. 
when SNe shells collide before $t_{shear}$. 
The reason is that the SNa shell radius at the time of shell collision 
($t_{coll}\propto f_{sn}^{-0.62}$) is $R_s\propto f_{sn}^{-0.19}$ and
the shell mass is $M_s\propto R_s^3\propto f_{sn}^{-0.57}$; therefore, 
the accretion rate $\dot M_{acc} \sim M_s(R_s/r)^2 f_{sn} \propto 
f_{sn}^{0.05}$. For the case of $t_{shear}/t_{coll}> 1$, the accretion 
rate is given by:
\begin{equation}
\dot M_{acc} \approx 0.3 \, n_3^{0.25} f_{sn}^{0.05} E_{51}^{0.7} 
     r_3^{-0.14} \; M_\odot\,{\rm yr}^{-1}.
\label{acc4}
\end{equation}

For $\dot M_{acc}$ to be independent of $r$ the density should scale
as $f_{sn}(r)^{-0.2}r^{0.5}$. However, for a non-equilibrium situation in
the early phases of galaxy formation and frequent mergers this equilibrium
density scaling is not applicable.
When a large quantity of gas is deposited within a few kpc following a merger
event or a tidal encounter, the high rate
of star formation and SNe at $r\sim 1$kpc would cause a rapid rate of 
accretion of gas to smaller radii, $\dot M_{acc}\sim M_s(t_{shear}) 
f_{sn}\sim 10^7$M$_\odot$ $f_{sn}$ yr$^{-1}$ after a lag of $\sim 10^7$ yrs,
which will continue until star formation and SNa explosions at smaller radii  
start to inhibit this large accretion rate; The subsequent accretion
rate would settle down to the value given by equation (\ref{acc4}).

Note that the cumulative effect of SNe is to create a random velocity field 
in the disk, but that does not automatically ensure accretion. We must 
have an outward angular momentum transport in order for accretion to proceed. 
An interesting example is that of convective instability in a disk; 
Ryu \& Goodman (1992) have shown that disk-convection transports 
 angular momentum inward, and therefore the turbulent velocity field 
 associated with it does not lead to any accretion. 
Similarly, the random velocity field in a disk is large when the SNa rate
is high and yet because shells collide before they are significantly 
deformed the outward transport of angular momentum increases very weakly
with $f_{sn}$ (eq. \ref{acc4}).
The accretion rate in terms of an effective $\alpha$-viscosity, 
$\nu \equiv \alpha R_s(t_{coll}) V_s(t_{coll})$, in the limit that
$f_{sn} > f_{sn}^{coll}$ is given by
\begin{equation}
\dot M_{acc} \sim 2\pi m_p n H \nu \sim 1.2 \alpha (H/0.1r) n_3^{0.39}
      f_{sn}^{0.24} E_{51}^{0.54} r_3^{0.53}\; M_\odot {\rm yr}^{-1}.
\end{equation}
Comparing this with equation (\ref{acc4}) we see that $\alpha\sim1$
for $f_{sn}\sim f_{sn}^{coll}$ (given by eq. \ref{fsn}), and $\alpha$ is smaller for a larger SNa
rate, although the effective $\alpha$ is dependent on $r$.

If SNe remnants punch through the disk in the vertical direction, but are
still confined by the galactic potential, then some fraction of gas leaving 
the disk will eventually fall back onto the disk at a smaller radius and
contribute to the net accretion rate. A SNa is confined to the galaxy provided that
\begin{equation}
R_s(\min\{t_{coll},t_{shear}\}) < H \quad {\rm or} \quad 
   V_s(R_s=H) < V_{orb}.
\end{equation}
If many SNe shells collide and coalesce before they are dispersed they would 
form super-shells and if their velocity is sufficiently high they can escape
the galactic potential. Otherwise, these super-shells would also contribute
to transporting gas to smaller radius.

\section{Star and bulge formation and growth of a central black hole}

The possibility of SNa-induced star formation has been discussed in numerous contexts, from the early 
universe to the present day Milky Way (e.g. Woodward 1976; Bedogni \& 
Woodward 1990; Yamada \& Nishi 1998; Mackey et al. 2003; Bratsolis et al. 2004; Joung \& Mac Low 2006; 
Johnson \& Bromm 2006; Sakuma \& Susa 2009; Le\~{a}o et al. 2009; Nagakura et al. 2009).  
Here we consider how this process may compete with the fueling of black holes
and contribute to the formation of galactic bulges.

For a Miller-Scalo initial-mass function (IMF) for stars (Scalo, 1986):
\begin{equation}
{dN\over dM} = 4.5 N_*\times \left\{
\begin{array}{lll} 
\hskip -5pt 1.9 (M/0.01M_\odot)^{-\alpha_1}   & \quad 0.01M_\odot< & M \le
    0.08M_\odot \\
\hskip -5pt (M/0.08M_\odot)^{-\alpha_2}   & \quad 0.08M_\odot< & M \le 0.5M_\odot \\
\hskip -5pt 6.25^{-\alpha_2} (M/0.5M_\odot)^{-\alpha_3}  &  & M>0.5M_\odot,
\end{array} \right.  \label{imf}
\end{equation}
with $\alpha_1=0.3$, $\alpha_2=1.8\pm0.5$, $\alpha_3=2.3\pm0.7$ (the 
parameters are taken from Kroupa, 2001), and $N_*=\int dM \, dN/dM$.
The mass fraction in high mass stars ($M_*\gta 8 M_\odot$), 
capable of SNa explosion, to the total star mass is about 0.3, and the 
fraction by number is about $4\times10^{-3}$. Thus the expected SNa 
rate is 0.3 yr$^{-1}$ for the Miller-Scalo IMF and a star formation rate (SFR) of 
10 M$_\odot$ yr$^{-1}$, which is of the order observed in ULIGals and
needed for forming galactic bulges in L$_*$ galaxies.  

Indeed, this is consistent with the following simple estimate of
the star formation rate in the gaseous disk:

\begin{equation}
SFR \sim f_{*} M_{\rm disk} / t_{\rm ff},
\end{equation}
where $M_{\rm disk}$ is the disk gas mass, $t_{\rm ff}$ is the free-fall 
time, and $f_{*}$ is the efficiency with which gas is turned into stars 
on a free-fall time,
taken to be of the order of $f_{*}\sim$ 0.01 (e.g. Krumholz \& Tan 2007).  Then,
for a disk of 10$^8$ - 10$^9$ $M_{\odot}$, the SFR is about 1-10 $M_{\odot}$ per
year, assuming a density of 10$^3$ cm$^{-3}$. 

If a similar fraction ($\sim$ 1\%) of gas in SNe shells is turned into stars (before 
shells collide) then the resulting star formation rate would be
$\sim 140 M_\odot n_3^{0.54} E_{51}^{0.4} r_3^{1.1} f_{sn}^{0.4}$ yr$^{-1}$, 
and that would result in a SNa rate of $\sim 5 n_3^{0.54} E_{51}^{0.4}
 r_3^{1.1} f_{sn}^{0.4}$ yr$^{-1}$ if the IMF for these ``daughter'' stars 
were given by equation (\ref{imf}). However, we show below that formation 
of massive stars -- those capable of SNa explosion -- is suppressed in SNe
remnants when $f_{sn}\gta 0.5$.

Star formation in a SNa shell is suppressed on large length scales due to 
the transverse relative velocity gradient in the remnant. 
We calculate this length scale as well as the Jean's length, and estimate 
the rate of star formation in SNe-remnants.

The velocity field in a SNa remnant seen by a comoving observer in her 
neighborhood is: $\vec{\delta V_s} \sim (V_s/R_s) \vec{\delta r}$; where
$\vec{\delta r}$ is the position vector, tangential to the shock front, 
pointing from the comoving observer
to a point in her neighborhood. For a gas clump to be able to 
collapse, the differential velocity across the clump, $|\vec{\delta V_s}|$,
 should be smaller than the gravitational escape speed, i.e.
for a clump of size $\ell$, $|\vec{\delta V_s}|\sim
V_s(\ell/R_s)< [G M_s(\ell/R_s)^2/4\ell]^{1/2}$ or
\begin{equation}
\ell \lta {G M_s\over 4 V_s^2}, \quad \quad \quad M_\ell \sim M_s 
  (\ell/2R_s)^2 \sim {G^2 M_s^3\over 64 V_s^4 R_s^2}
\end{equation}
Using equations (\ref{rsnow})--(\ref{msnow}) we find 
\begin{equation}
  M_\ell \sim 9\times10^{-12}\left[{t\over t_{snow}}\right]^{4.9} 
    n_3^{-{6\over17}} E_{51}^{{31\over17}}\;M_\odot,
\end{equation}
We see from equation (\ref{vsnow}) that at $t/t_{snow}\approx 137 
n_3^{0.17} E_{51}^{0.09}$ the shell speed is $V_s=20$ km s$^{-1}$,  
and at that time $M_\ell \sim 0.2 n_3^{0.48} E_{51}^{2.2}$ M$_\odot$;
the maximum star mass scales as $V_s^{-7.1}$. 

Star formation is 
disrupted when shells collide\footnote{Collision between SNe shells 
disrupts star formation because of turbulence generated in these collisions.
Once turbulence dies out -- in about a shock crossing time -- star 
 formation can resume provided that the merged shell is not hit again by
a high speed shock.}.
The average speed of SNe shells when they collide, at time $t_{coll}$, is 
$V_s\sim$ 14 km s$^{-1} f_{sn}^{0.4} r_3^{-0.8} n_3^{-0.5} E_{53}^{0.4}$
(obtained from equations 6 \& \ref{tcoll}).  
For a small SNa rate, $f_{sn}< 0.5$ yr$^{-1}$, $V_s$ can drop 
down to a value where massive stars capable of SNa explosion can form 
before SNe shells collide and star formation is disrupted; however, 
since lower mass stars form first in SNe remnants they can perhaps
significantly suppress the formation of more massive stars.
At a higher SNa rate, formation of stars more massive that a few
solar mass is suppressed; massive stars could still form in the disk 
after the turbulence generated by shell collisions has subsided and gas
has cooled down.
This suggests that SNe explosions in a gaseous disk might occur in waves 
of high and low activity, and during periods of high activity Miller-Scalo
IMF is truncated above a few solar mass, and even during periods of lower
SNa rate the IMF could be more bottom-heavy than the standard IMF.

\subsection{The thermal state of the pre-shock gas and the shock-induced stellar IMF}

A firm lower limit to the shock front speed, which in turn sets an upper 
limit to the masses of the stars which may form behind the shock front, is 
set by the sound speed of the medium upstream of the shock front when 
SNa shock weakens and turns into a sound wave.  In order to calculate 
the sound speed of the upstream gas, we must determine the thermal state 
of this gas, and we discuss two possible cases in this subsection. 

\subsubsection{High supernova rate}

For the first case, we consider a galactic disk with a high star 
formation rate (SFR), in which the dominant 
process affecting the thermal state of the gas is photo-heating of 
the gas by massive stars which eventually explode as SNe.  Considering that 
these stars will each live for roughly 10$^7$ yr, 
the average distance between these massive stars within radius $r$ of the disk,
 following equation (18), is 
\begin{equation}
d_{sn} \approx 0.6\, \frac{r}{1 {\rm kpc}}\, f_{sn}(r)^{-1/2}\;{\rm pc}.
\end{equation}
The number of ionizing photons emitted per second by 
such massive stars is $\ga$ 10$^{49}$ (e.g. Osterbrock \& Ferland 2006).
 Taking the average gas density to be $\sim$ 10$^3$ cm$^{-3}$, we find 
the radius of the Str\"{o}mgren spheres surrounding massive stars 
(Str\"{o}mgren 1939) to be of order $\sim$ 1 pc. Thus, we expect that for an
average SNa rates of $f_{sn}\gta$ 0.4 yr$^{-1}$ the H~II regions surrounding 
the massive stars in the disk will overlap, and therefore SNe shocks will 
propagate into such photoionized regions.  Largely 
independent of the metallicity of the gas, the temperature in such 
a photoionized region will be of the order of 10$^4$ K, and the sound 
speed $\sim$ 10 km s$^{-1}$. Therefore, SNe shocks in this case will 
generally dissipate and turn into pressure waves once they have slowed to 
speeds of $\sim$ 10 km s$^{-1}$.  This suggests, following equation (29), 
that the IMF of stars formed in the material swept up in SNa shocks is 
likely cut-off at a few solar masses.

\subsubsection{Low supernova rate}

We next consider the case of a lower SFR, corresponding to $f_{sn}\lta$ 
0.4 yr$^{-1}$, for which the distance between massive stars is greater 
than their average Str\"{o}mgren radii. In this case, SNa shocks will 
generally propagate into so-called relic H~II regions in which gas 
that was previously photoionized by the progenitor star is recombining 
and cooling; equations (\ref{rsnow}) \& (\ref{vsnow}) show that, in general, $R_s\gta 1$pc 
(Str\"{o}mgren radius) when the SNa shell speed has dropped to $\sim10$km
s$^{-1}$. However, even for larger Str\"{o}mgren radii, 
the upstream relic H~II region gas can in some cases
cool to temperatures well below 10$^4$K, which allows for the formation 
of stars with masses greater than a few solar masses.      

The results of a numerical calculation of relic H~II region gas temperature 
ahead of the shock are shown in Fig. 1. This calculation assumes that the 
density of the gas remains constant at the fiducial value of 
n = 10$^3$ cm$^{-3}$, and that the gas cools only radiatively through 
atomic transitions of metals, which is a reasonable assumption for the 
case we consider here (but see Jappsen et al. 2009, for the case of star 
formation in a more isolated environment). In principle, the radiation 
emitted  by the shocked gas can send a radiative precursor ahead of the 
shock and heat the upstream gas, however, we neglect this effect as at 
the shock velocities at which the shocks stall, i.e. $\la$ 20 km s$^{-1}$, 
radiative precursors do little to ionize or heat the upstream gas 
(Shull \& McKee, 1979).

Figure 1 shows results for four different metallicities: 10$^{-1}$, 
10$^{-2}$, 10$^{-3}$, and 10$^{-4}$ $Z_{\odot}$. The metallicity-dependent
cooling rate of the gas is taken from Mashchenko et al. (2008), who 
provide a fitting formula to the cooling function calculated by 
Bromm et al. (2001). The left panel of Fig. 1 shows the temperature of 
the upstream gas as a function of the time from the death of the 
central star. The right panel shows the square of the Mach number for the
shock front. The SNa shock will stall once this ratio
approaches unity -- at this point the shock will turn into a pressure wave.
 
\begin{figure}[h!]
\begin{center}
\includegraphics[width=15.7cm,height=13.4cm]{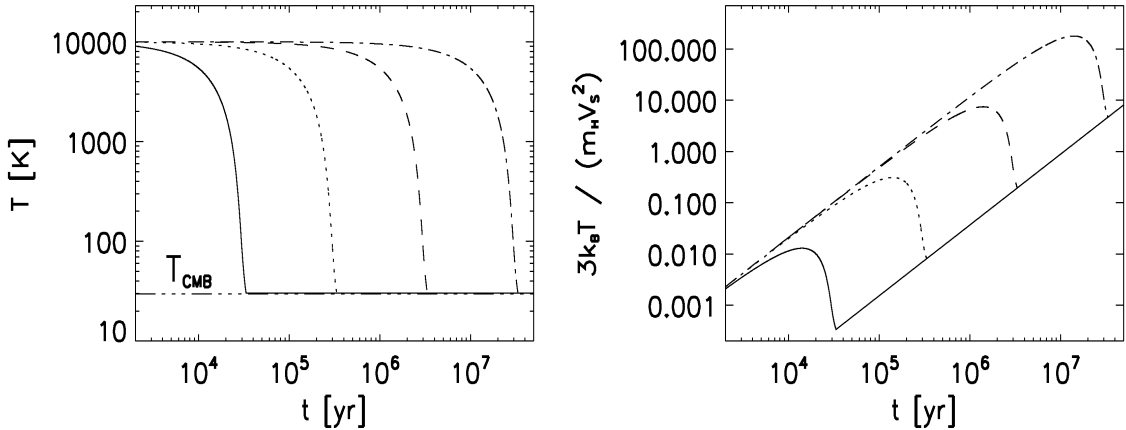}
\vspace*{-7cm}
\caption{The temperature of the medium upstream from the shock front (left
panel) and the ratio of the squares of the upstream sound speed and shock 
front speed during the snowplow phase (right panel), in a cooling relic 
H~II region with gas density of 10$^3$ cm$^{-3}$.  The shock wave 
turns into a compression wave when the ratio drops to $\sim 1$, which 
occurs at $t$ $\ga$ 10$^5$ yr for each of the gas metallicities considered 
here: the curves from left to right correspond to metallicities of 
10$^{-1}$, 10$^{-2}$, 10$^{-3}$, and 10$^{-4}$ $Z_{\odot}$, respectively. 
Note, however, that for lower upstream gas densities this occurs at earlier 
times, due to the lower cooling rate of the relic H~II region gas.   }
\label{fig1}
\end{center}
\end{figure}

For metallicities $\lta10^{-2}$ Z$_{\odot}$ the SNa shock stalls at 
a time $t_{stall}$ $\sim$ 10$^5$ yr, when the shock velocity is still 
$\sim$ 10 km s$^{-1}$. However, for higher metallicities the shock 
stalls at later times when $V_s$ has dropped to a smaller value and 
thus the maximum mass of SNa-induced stars can become $M_l\gta$ 5 
M$_{\odot}$ for shocks that last for $\ga$ 2 $\times$ 10$^5$ yr 
(see eq. 30). We note also that for the case we consider here of 
$f_{sn}$ $\la$ 0.4, $t_{coll}$ can easily exceed $t_{stall}$,
and therefore shell collisions will not interfere with SNa-induced 
formation of these more massive stars at late times.

\subsection{Stars, bulge and black hole}


As shown in the last Section, for $f_{sn}\gta$ 0.4 yr$^{-1}$, stars forming 
in SNe remnants 
have peculiar velocities of order 10--20 km s$^{-1}$ in radial direction 
with respect to the center of the explosion. When viewed from the 
galactic center the velocities of these newly formed stars in 
SNa remnants would have a random velocity dispersion of 
$\sim10$--20 km s$^{-1}$, and therefore these stars would tend to
form a bulge at the center. The velocity dispersion of the newly formed 
stars would increase with time as these stars are subjected to the 
stochastic gravitational field of the SNa remnant network. 
Moreover, these stars will suffer some hydrodynamical 
drag on their way out of the gaseous disk, and also will be subject to 
gravitational drag that will modify their velocity dispersion (eg. 
Artymowicz, 1994; Nayakshin \& Cuadra, 2005); the former is likely a
small effect due to the small cross-section for star-gas interaction,
but the latter can be a significant effect and needs to be included
in the calculation to determine the true random velocity distribution of 
stars formed out of SNe remnants.

The supernova led accretion would also deposit gas in the central parsec 
region of the galaxy at the rate given by equation (\ref{acc4}); at distances
smaller than $\sim1$pc the magneto-rotational
viscosity (Balbus and Hawley, 1991) is expected to be effective in 
transporting gas to the black hole at the center. We note that MRI might 
also operate at larger radius due to heating and ionization produced by SNe. 

Let us assume that a fraction $f_*$ of supernova remnant mass is converted 
to stars before shells collide with another shell.
 Using equations (\ref{msnow}) and (\ref{tcoll}) 
we estimate the star formation rate in SNa remnants to be
\begin{equation}
\dot{M_*} \approx (1.5\times10^4 M_\odot)\, f_*\, f_{sn}^{0.4} n_3^{0.54}
    E_{51}^{0.4} r_3^{1.1} \; {\rm yr}^{-1}
\label{mdotstar}
\end{equation}
Therefore, the ratio of star formation \& accretion rates for $f_{sn}\gta
0.5$yr$^{-1}$ is $\sim$5x$10^4 f_* f_{sn}^{0.3} n_3^{0.3} E_{51}^{-0.3}
 r_3^{1.2}$ (eqs. \ref{acc4} \& \ref{mdotstar}). For $f_*\sim 10^{-2}$ 
and $f_{sn}\sim 1$ yr$^{-1}$ this ratio is
of order a few hundred. In the case of low SNa rate (when the remnant
survives until its velocity drops to $\sim10$ km s$^{-1}$ and then
turns into a compression wave) the ratio is $3\times10^4 f_*
n_3^{0.7} E_{51}^{-0.7} r_3^{2}$, which is also of order a few hundred
for $f_*\sim10^{-2}$; this last expression was obtained by using eqs. \ref{rsnow}
\& \ref{msnow} for $R_s$ \& $M_s$ corresponding to shell velocity of 10 km s$^{-1}$
(eq. \ref{vsnow}) to calculate star formation and accretion rate (eq. \ref{acc2}).

This ratio of star formation rate to accretion rate is similar to
 the reported ratio of bulge and BH mass in galaxies (Gebhardt et al. 2000;
Ferrarese \& Merritt 2000), and the scenario we have described offers a 
plausible physical explanation for this correlation.
We note that a number of well known feed back processes have been
left out in the calculations presented in this paper (cf. Cattaneo et al.
2009), and these might significantly modify star formation and accretion rates.

\section{Summary}

We have analyzed the effect of supernovae occurring within the central kpc
region of a gaseous disk on the formation of stars and transport of gas
from $\sim 1$kpc to a few pc of the galactic center. The outward transport of 
angular momentum facilitated by SNe explosions allows for the inward transport 
of gas that feeds the central black hole. This is a process
which may take place quite generically in any galactic disk hosting star 
formation, although it may not be the dominant process affecting 
black hole accretion.

We have shown that associated with the inward transport of gas swept up by SNe 
is the shock-induced formation of stars which are born with a random peculiar
velocity of $\sim 10$ km s$^{-1}$. This velocity dispersion increases 
with time as a result of the stochastic gravitational field associated with 
filamentary SNa remnants; The stars formed in SNa remnants contribute
to the stellar population of a central bulge.

Due to the divergent velocity field of an expanding SNa shell there is a 
maximum length scale for fragmentation of shells or an upper limit to the
mass of stars formed in SNa remnants; the SNa-induced stellar IMF is cut-off
above a few solar masses and more massive stars can only form if and when a
SNa shell slows down to $\sim10$ km s$^{-1}$.

We note that numerous observations have suggested connections, such as 
we have considered in the present work, between star formation, black hole 
accretion, and the formation of a stellar bulge. 
Heckman et al. (2004) suggest that star formation and black hole accretion rates
are correlated, and Chen et al. (2009) report an empirical relation between 
supernova rate and gas accretion rate on the central black hole (see also 
Xu \& Wu 2007). Furthermore, Page et al. (2001) report observations 
suggesting that central black holes and stellar spheroids form 
concurrently, and Genzel et al. (2006) describe observations of 
a galaxy hosting both an accreting black hole and a central stellar 
bulge, with no evidence of a major merger. 
Hydrodynamic simulations have demonstrated that SNa feedback may produce 
spherical distributions of stars in dwarf galaxies (Stinson et al. 2009)
and in the inner portions of the Galactic bulge (Nakasato \& Nomoto 2003).   
We would like to point out the recent work of Wang et al. (2009) that models
SNe induced turbulence as an effective viscosity and describes the evolution 
of a gaseous disk.

A limitation of this work is that we have ignored radiative feedback effects 
which are known to control the steady state accretion
rate onto the black hole (e.g. Ostriker et al. 1976, Proga et al 2008, 
Milosavljevic et al. 2008) and probably also affect the formation rate
of stars in the central kpc.  Ultimately, large scale
simulations resolving the long term evolution of individual SNe remnants, 
star formation, the multiphase ISM, and the feedback effects of accretion onto 
a central black hole will be required to more fully elucidate what 
role SNe-induced accretion and star formation play in galaxy formation.

\section{Acknowledgments}
JLJ gratefully acknowledges the support of a Wendell Gordon Fellowship from the University 
of Texas at Austin. 
We thank the referee for many constructive comments that helped improve
the presentation significantly.

\end{document}